\documentclass{article}
\usepackage{graphicx}
\usepackage{etoolbox}
\AtBeginEnvironment{quote}{\par\small}

\usepackage[superscript,biblabel]{cite}
\usepackage{url}
\title{A Multi-Scale Cognitive Interaction Model of Instrument Operations at the Linac Coherent Light Source}
\usepackage{authblk}

\author[1]{Jonathan Segal}
\affil[1]{\footnotesize{School of Information Science, Cornell University. jis62@cornell.com}}
\author[2]{Wan-Lin Hu}
\affil[2]{Energy Sciences, SLAC National Accelerator Laboratory. wanlinhu@gmail.com}
\author[3]{Paul H. Fuoss}
\affil[3]{Linac Coherent Light Source, SLAC National Accelerator Laboratory. fuoss@slac.stanford.edu}
\author[4]{Frank E. Ritter}
\affil[4]{College of IST, Penn State. frank.ritter@psu.edu}
\author[5,*]{Jeff Shrager}
\affil[5]{Chief Scientist, Bennu Climate, Inc. and Adjunct Professor, Symbolic Systems Program, Stanford University. jshrager@stanford.edu}
\affil[*]{Corresponding author}

\date{December 2024}

\begin{document}

\maketitle

\begin{abstract}
The Linac Coherent Light Source (LCLS) is the world’s first x-ray free electron laser. It is a scientific user facility operated by the SLAC National Accelerator Laboratory, at Stanford, for the U.S. Department of Energy. As beam time at LCLS is extremely valuable and limited, experimental efficiency --- getting the most high quality data in the least time --- is critical. Our overall project employs cognitive engineering methodologies with the goal of improving experimental efficiency and increasing scientific productivity at LCLS by refining experimental interfaces and workflows, simplifying tasks, reducing errors, and improving operator safety and stress. Here we describe a multi-agent, multi-scale computational cognitive interaction model of instrument operations at LCLS. Our model simulates aspects of human cognition at multiple cognitive and temporal scales, ranging from seconds to hours, and among agents playing multiple roles, including instrument operator, real time data analyst, and experiment manager. The model can roughly predict impacts stemming from proposed changes to operational interfaces and workflows. Example results demonstrate the model's potential in guiding modifications to improve operational efficiency. We discuss the implications of our effort for cognitive engineering in complex experimental settings, and outline future directions for research. The model is open source and supplementary videos provide extensive detail.  
\end{abstract}

\section{Introduction}

LCLS, the Linac Coherent Light Source, located at the SLAC National Accelerator Laboratory, is a world-leading x-ray source, hosting hundreds of high energy x-ray experiments annually. These experiments probe fundamental physics as well as the nano-scale, high speed physics underlying fields as diverse as biology, chemistry, and materials science.\cite{lclsoverview} LCLS beam time is a limited resource that must be efficiently utilized. Our goal is to apply cognitive engineering methods to optimize LCLS user interfaces (UIs) and workflows, that is the way that scientists and engineers gather information, reason, and take action with regard to the system.  

Cognitive Engineering is a methodology tailored to analyze and improve ``human-in-the-loop'' settings --- those where humans interact with complex engineered systems.\cite{WilsonEtAl2013,nrc2007}  The method combines qualitative and quantitative experiments, mathematical analysis, and computer simulations, with the goals of improving efficiency by making tasks simpler and less error prone, improving participants' safety, and reducing their stress. 

By creating a computational cognitive model of some aspects of the UI and workflow of LCLS operations we can predict the impact of proposed changes before they are actually implemented. This approach is not unique to LCLS. Just as cognitive engineering applies broadly, computational modeling is equally broadly applicable. However, modeling is more important in cases such as LCLS because of the unique challenge posed by scientific user facilities. Due to its complexity, cost, and limited beam time, it is not feasible to run typical human-factors experiments, nor to create a ``flight simulator'', at  LCLS. Moreover, LCLS experiments present an ever-changing panoply of experiments, nearly all of which are at least somewhat unique, and so do not have the stable regularity of, say, aviation or power plant operations. Computational modeling should be especially helpful in such settings as changes can be explored without running live human experiments.\cite{nrc2007} 

In this paper we describe the LCLS setting, and a computational model that can make rough predictions about the impact of proposed changes to interfaces and workflows. A unique feature of our model is that it integrates simulations at multiple scales, and models aspects of individual cognition, teamwork, planning, and decision-making both in-the-moment and across multiple data-taking runs (measurements). Although the model represents these only roughly, it is able to make sensible predictions regarding the impact of a range of proposed changes. Because of the cost and limited availability of beam time, even rough predictions may be useful in increasing the efficiency of LCLS operations. 

We begin with a rich description of the LCLS setting so that the reader can understand the rationale behind the model's design. Next we describe the model and demonstrate its basic capabilities, concluding with future opportunities for this approach. The model is open source.\cite{githubmodel} Demos and code walk-throughs, are provided in three supplementary videos.\cite{vids} 

\pagebreak
\section{Setting: The Machine and Instruments}

The LCLS system involves an enormous number of physical and software components, operated by multiple teams of scientists and engineers. The 2 kilometer-long linear accelerator (hereafter ``linac'') and a 200 meter long magnetic ``undulator'' form an x-ray free electron laser (XFEL), often referred to together as ``the machine''. The machine produces 120 extremely bright 50 femtosecond x-ray pulses every second, referred to as “the beam”. After propagating through various diagnostics and beam shaping optics, the beam arrives at an ``instrument'' whose purpose is to collect data about what happens when the beam interacts with materials of interest, generally called ``targets'' or ``samples''. (The target is a sample that is actually in the beam.) There are several of these room-sized instruments, although only one is usually involved in a particular experiment.\cite{Traweek1}

The machine and instruments are operated from separate control rooms. The control room operating the machine is called the Accelerator Control Room (ACR). Each instrument has its own Instrument Control Room (ICR, Figure 1), located near its hutch at the target-end of the system. Here we are mostly concerned with activities in the local control room of a specific instrument, called CXI, which is specialized for x-ray crystallography.\cite{xcipage} 

\begin{figure}[!htb]
    \centering
      \makebox[\textwidth]{\includegraphics[width=\textwidth]{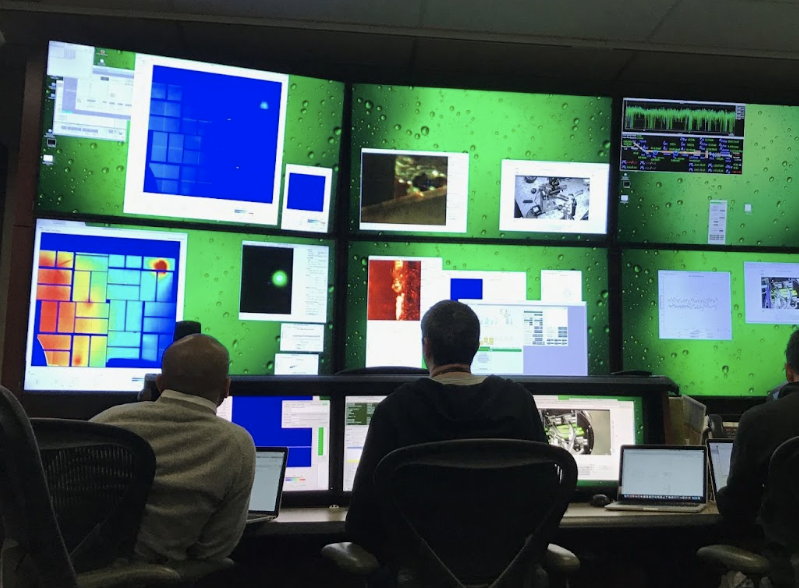}}
        \caption{Typical view of an Instrument Control Room. (Photo by the authors)}
    \label{fig1}
\end{figure}

\begin{figure}[!htb]
    \centering
      \makebox[\textwidth]{\includegraphics[width=\textwidth]{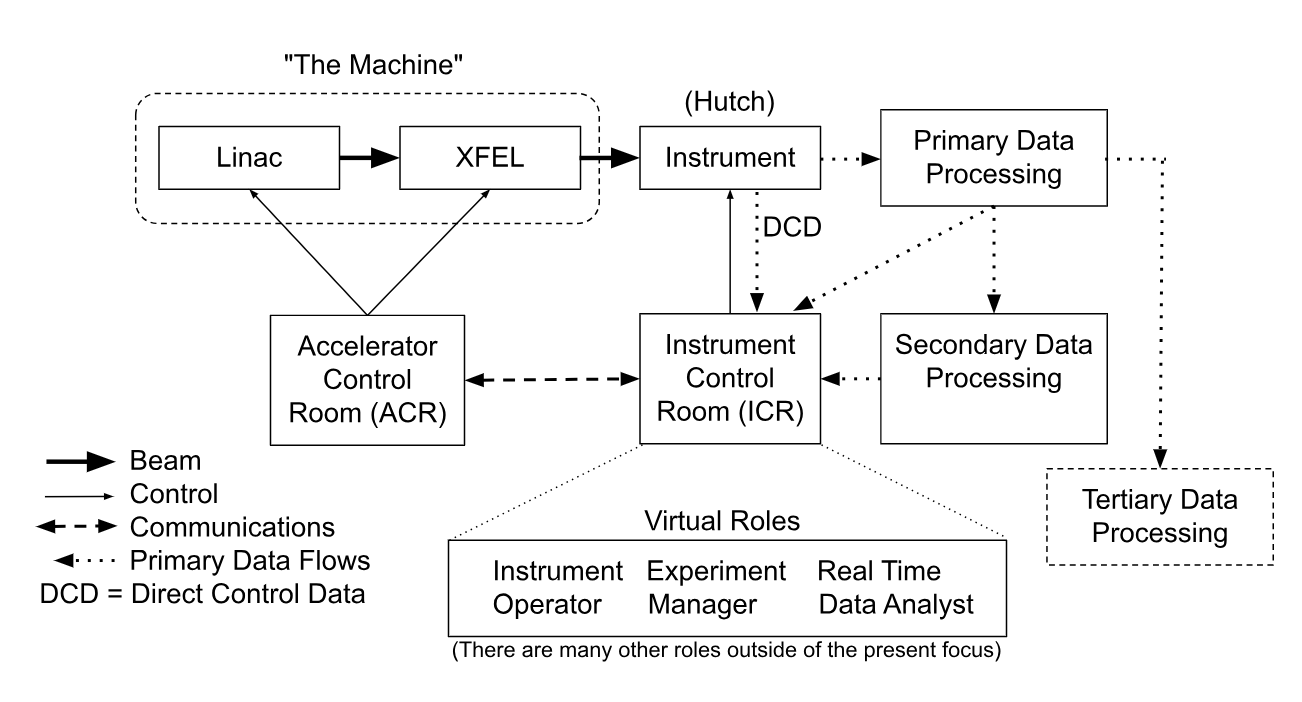}}
        \caption{The machine (linac + XFEL) delivers x-ray pulses to an instrument where the x-rays strike a target producing data which is processed by an analytic pipeline. The system is operated through a close and continuous collaboration between scientists and engineers in the ACR and ICR who play various roles.}
    \label{fig2}
\end{figure}

\newpage
 \ref{fig2} abstractly depicts the layout of subsystems involved in SLAC/LCLS operations, and their typical interactions (see also Ritter, 2022\cite{ritter2022}).

LCLS experiments depend upon the x-ray beam created in the machine, which is ``handed off'' to the instrument. Because the success of the experiment depends upon creation of x-ray pulses with the desired characteristics at the machine end, a constant interaction takes place between the instrument operators in the ICR, and the machine operators in the ACR. As the ACR and the ICR are about a 15 minute walk, or a 3 minute drive, from one another, this interaction takes place mostly by phone or through screen sharing. In the present work we make the unrealistic simplification that beam delivery --- the machine end of the process --- is unproblematic, and focus entirely on operations at the instrument end of the system.

\clearpage
\section{Experiments and Measurements} \label{expmes}

\begin{figure}[!htbp]
    \centering
      \makebox[\textwidth]{\includegraphics[width=\textwidth]{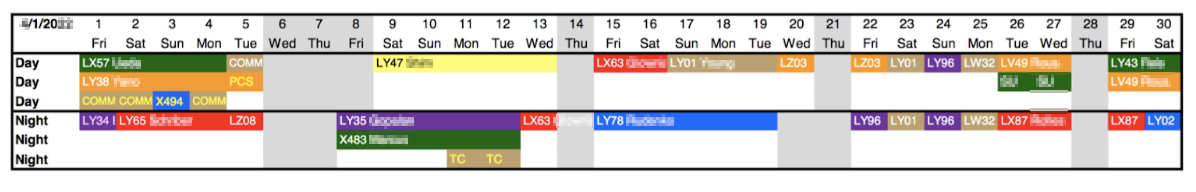}}
        \caption{An example LCLS schedule.\cite{LCLSSchedule} Reprinted with permission from SLAC.}
    \label{fig3}
\end{figure}

An ``experiment'' is a contiguous period, usually 3-5 days, during which users are actively engaged with the system. Figure \ref{fig3} shows a typical schedule for one month at LCLS. The teams using the beam are color coded by hutch (red is CXI). Grey blocks represent scheduled downtime, e.g., for maintenance. 

Experiments have many sub-steps. We are primarily concerned with ``measurements'' ---  contiguous periods during which the x-ray beam is intended to be striking a target or series of targets, and the team is actively taking data. There are usually many dozens of measurements made over the several-day period of an experiment, generally making up the bulk of an experiment. Measurements can range from a few seconds (if something goes wrong or is only being briefly tested), to many minutes, as multiple samples are cycled into the path of the beam. As time is the most limiting resource, the time that measurements take will be of central concern in our model.

Most LCLS experiments utilize a variety of types of samples (aka. targets when in the beam), including ``dummy'' targets, used to align and calibrate the instrument, ``control'' targets that have known properties, and ``critical'' targets, the samples for which the experiment is designed to gather new data. Because the critical samples are often costly and of limited availability, the experiment is planned to maximize high quality data (data with high signal-to-noise ratio, SNR) collected on these. Although an enormous amount of careful planning goes into experiment design before the experimental team arrives at SLAC, unplanned issues always arise, so the team engages in continuous, on-site, real-time decision-making to maximize the collection of high quality data on  critical samples. These concepts will appear as ``Performance Quality'' and ``Target Error'' in the model (\S \ref{tepq}).

The primary decisions to be made are how long a given measurement should continue, and in what order different samples should undergo measurement. Because beam time is limited, gathering data on critical samples must begin as soon as possible. However, because issues often arise during the experiment's initial beam time, it is sensible to use dummy or control targets until it is clear that the instrument is stably producing high SNR data. As the experiment nears the end of its allocated beam time, the team may wish to get as much data as possible on critical samples, even if the system is not producing high quality data, and especially if the experiment did not go as planned, for example, due to linac down time. These trade-offs define the top layer of our model (\S  \ref{macro}).

\section{Experimental Teams and Virtual Roles}

The scientific productivity of LCLS depends on highly complex human-machine choreography enacted by combinations of internal and external scientists and engineers who we call ``experimental teams''\cite{Traweek88}. Although there is extensive planning for each experiment, the entire team usually only comes physically together for the few days during which measurements are being taken at LCLS. Team members often have diverse backgrounds (biologists, chemists, physicists, engineers, etc.), and diverse experience ranging from students who are at LCLS for the first time, to full-time LCLS staff such as the ``instrument scientists'' who develop and usually operate the instruments. We have observed experiments with as few as one, and as many as ten or more people in an ICR during a running measurement. There are also, commonly, remote participants, interacting by phone or web technologies.

\subsection{Operator, Analyst, and Experiment Manager}

The success of an experiment rides on the team's ability to rapidly develop an effective team workflow in order to conduct as many measurements as possible with high SNR on important samples. We identify three ``virtual roles'' (Figure \ref{fig2}) which are always represented, regardless of the number of separate people involved in an experiment: The \textbf{Instrument Operator}, the \textbf{Real Time Data Analyst}, and the \textbf{Experiment Manager}.

\begin{figure}[!htbp]
    \centering
      \makebox[\textwidth]{\includegraphics[width=0.75\textwidth]{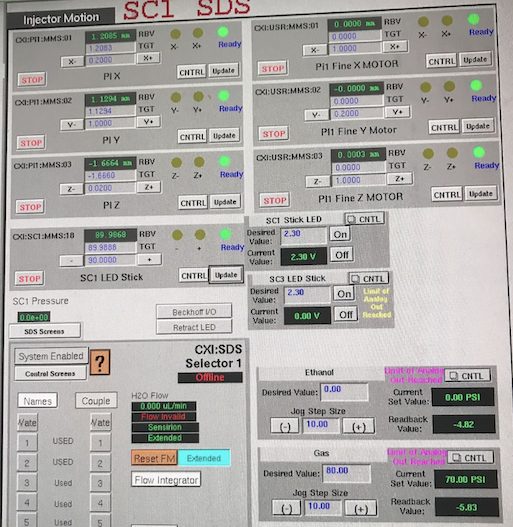}}
        \caption{A typical control window. Most of the depicted controls are motor controls like those in our model. In addition to using the left and right buttons (e.g., X+ X- Y+ Y-), the motor position can be changed by directly entering a new value in the text box. For each such control there is almost always a command-line equivalent. Which type of interaction is used in a particular setting depends on many factors, but especially the preferences of the operator. (Photo by the authors)}
    \label{fig4}
\end{figure}

The \textbf{Instrument Operator} actively controls the instrument. They are usually an individual, and are almost always seated at the screens displaying the status of, and permitting control of the hundreds of instrument settings, such as motor positions and valve and shutter settings (e.g., Figure \ref{fig4}). In making decisions the operator relies on information coming directly from primary data processing, as well as  from the other team members and the ACR.  

\begin{figure}[!htbp]
    \centering
      \makebox[\textwidth]{\includegraphics[width=\textwidth]{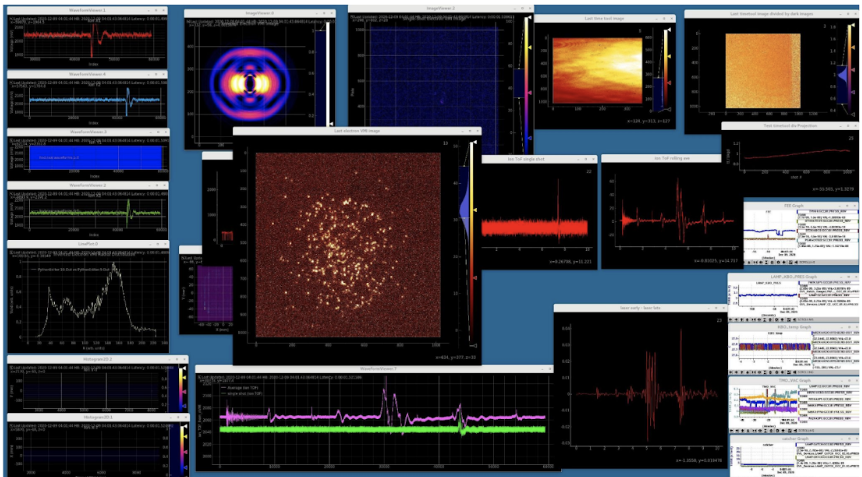}}
        \caption{Real time AMI plots.\cite{amifig} Reprinted with permission from SLAC.}
    \label{fig5}
\end{figure}

The \textbf{Real Time Data Analyst} analyzes the data flowing from the instrument into the data pipeline using tools such as the AMI (Analysis Monitoring Interface\cite{ami}, Figure \ref{fig5}). We call this ``Secondary Data Processing'', and its primary function is to estimate SNR of current targets and report this to the experiment manager.

The \textbf{Experiment Manager} is responsible for overall execution of the experiment. Although each experiment starts out with a specific experiment plan, dynamic (``opportunistic'') re-planning almost always takes place based upon many factors.\cite{Hayes-Roth1995} The experiment manager uses the information from the operator and data analyst to monitor the plan and adjust it as needed. The important decisions here are usually whether to continue or abort the current measurement, and what sample to measure next.  

\pagebreak
These three roles are often shared in complex ways across any number of individuals. For example, the instrument operator might also be the data analyst, and either of these might also be the experiment manager. With the exception of the instrument operator, the people filling the other two roles are sometimes involved by phone or screen sharing. How these roles are distributed has implications for the efficiency of communication and cognition. Our model assumes that each role is occupied by a separate individual, and we assume that the participants are highly expert in their roles.

Once the instrument is prepared and a target is in place, the instrument and machine operators work together to align the beam to reach the sample in the desired manner. They ensure that data is being taken correctly by the instrument, and that the various downstream data pipelines are working as intended. All of this involves hundreds of physical and computational adjustments. Once this setup is completed the experiment itself can begin. What follows is the series of measurements that comprise the bulk of the experiment, and which are our primary focus.

\pagebreak
\section{Cognitive Aspects of LCLS Experiment \label{resources} \linebreak\mbox{Operations}}

LCLS operations are an ideal setting to understand team-based, real-time, problem-solving, characterized by complex opportunistic interplay of numerous inter-related goals and the physical and cognitive activities that the team deploys to reach these goals.\cite{Graesser2018, Hutchins2010} 

In acting towards their goals the team utilizes ``resources'', the most important of which are actions and their affordances (together, ``afforded actions'').\cite{Norman1988} One can think of an afforded action as one that can be accomplished through manipulation of some sort of control, for example moving a mirror by virtue of stepping a motor, which is afforded by the motor controls (e.g., Figure \ref{fig4}). Afforded actions can include those related to ``information resources'', that is displays, graphics and cameras (e.g., Figure \ref{fig5}), auditory signals, paper documents, electronic logs, and even tactile feedback from which a participant may gather information. (These ``afford'' the ``action'' of obtaining certain types of information.) Of course, information resources often require their users to take action in order to obtain the desired information, and information can be communicated as well as received. To highlight one interesting example, Post-Its are quite commonly stuck to various parts of instruments, controls, and screens to facilitate the remembering and efficient communication of important information. And an important resource/artifact that affords both information-gathering and control actions is the experiment plan itself, which can be queried regarding predicted times, and acted upon (controlled) by modification. 

People can also be thought of as offering affordances. For example, the instrument operator's expertise affords the experiment manager actions such as getting data and terminating the measurement, and the expertise of the linac operators in the ACR affords  actions as complex as requesting various adjustments to the beam, or as simple as finding out when the machine is likely to return from unexpected down-time. 

Finally, two important non-physical resources, time and attention, play a critical role in this setting, as they do in almost all problem-solving settings. The operating team must attend and respond to a very large number of information resources in order for a measurement to succeed. There are often dozens of windows open across numerous screens, including personal laptops. The operator interacts with these partly ``demandantly''\cite{Demandance}  --- for example responding to alarms --- and partly proactively, that is, with the intention of making a change that is \textit{not} in response to an alarm. Both time and attention play a large role in our present model, and will reappear as ``Functional Operability'' and ``Functional Acuity'' (not respectively) in the discussion of the model in section \ref{peak-chasing}.

\subsection{The Central Decisions: What to Measure, in What Order, and for How Long}

Once the instrument is producing high SNR data on dummy or control samples, the heart of the experiment can begin by taking measurements from critical samples --- those from which real data is desired. As mentioned above, each experiment is planned in great detail, yet if enough problems arise the team might run out of beam time or critical samples. Importantly, because of the amount of data involved, and the limited time available to gather and analyze it, final (tertiary) data analysis usually cannot be accomplished in real time. Therefore, the team can only \textit{roughly} tell, through secondary data analysis, whether they are gathering satisfactory data on the current sample. Thus the decisions of what to measure when, and for how long are the experiment manager's most critical decisions. These are made, of course, in consultation with the rest of the team, often including the machine operators in the ACR. 

One would think that it would make the most sense to carry out measurements on the more important sample as early as possible because if problems arise that lead to lower SNR, one can use more time to take more data on critical samples. However, there may be limited sample material, especially of the most important samples, and early problems are likely to have been resolved, improving SNR. These considerations vie in the opposite direction, suggesting that if things are not working well, but it is still early in the overall experiment period, it may be better to hold back precious critical samples hoping that the problems resolve. But how long should one wait? And if the data is marginal, and getting worse, should one continue a measurement to wring as much out of the current sample as possible, or abort it and move on to more important samples, given an impending hard stop? And once the ongoing measurement is completed, which of the remaining samples should be measured next? Perhaps a previous less-than-satisfactory measurement should be repeated? 

Because of the complexity and importance of making these decisions as correctly as possible, we focus on these in our model. 

\section{A Multi-Scale Cognitive Interaction Model of LCLS Instrument Operations}

Our goal is to predict the potential impacts of UI and workflow changes in the most impactful aspects of LCLS operations. Because of the difficulty of carrying out direct user studies, or of employing a ``flight simulator'' approach in the LCLS setting, we chose to create a computer simulation (model) of some of the most important human-in-the-loop aspects of experiment operations, in particular of the phenomena relevant to the core decisions of when and whether to continue or abort a measurement. Our goal is to use this model to predict the direction and rough shape of changes in experiment efficiency likely to result from changes in instrument interfaces, controls, and procedures. 

In what follows we discuss the central concepts represented in our model. The three video supplements provide much more detail.\cite{vids} The first is a conceptual overview, the second demonstrates the model, including discussion of the results depicted in Figures \ref{fig7}-\ref{fig9}, and the third reviews the code, which is available on github.\cite{githubmodel}

\subsection{Multi-Agent, Multi-Scale}

Real LCLS experimental decision-making takes place in a very complex setting where many extrinsic factors are involved, such as which samples are available, what the value is of those samples, the stability and performance of all aspects of the system, how far through the experiment they are, and how much time the remaining high-value samples are likely to take, as well as large-scale factors such as machine down-time and the requirement that team members occasionally sleep. Our model, of course, represents a highly simplified version of these factors.

\begin{figure}[!htbp]
    \centering
      \makebox[\textwidth]{\includegraphics[width=\textwidth]{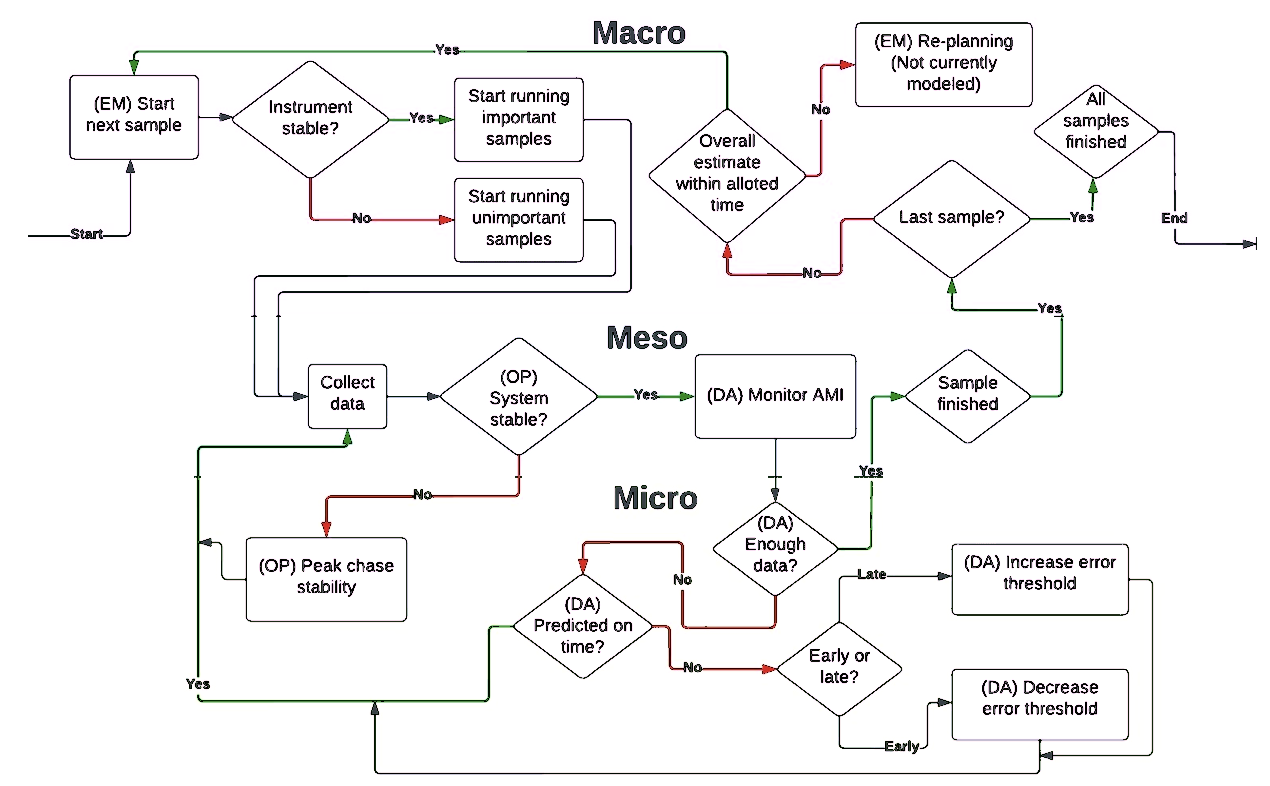}}
        \caption{Information flow and inter-scale interaction during LCLS operations. ``Micro'' reasoning and action is local in time, relating to the moment-to-moment stability of the system required to get any data at all as well as high quality data; ``Meso'' relates to a larger time scale, over multiple data points; and ``Macro'' is at the scale of the entire experiment, over multiple measurements.}
    \label{fig6}
\end{figure}

We explicitly model the three primary agents: The Operator is directly in control of the instrument. The Real Time Data Analyst determines when the data is good enough to stop the run, and provides updates to the Experiment Manager who is responsible for the overall control of the experiment and communicates with the entire team to understand the state of the experiment and to make decisions that improve the quality of the data and the efficiency with which that data is generated. These roles implicitly create three temporal scales, as depicted in Figure \ref{fig6}. 

\begin{itemize}
     
\item[]Phenomena at the \textbf{Micro-Cognitive scale} relate to the details of controls and displays. Here we see the reasoning, decision-making, and skilled performance that we collectively call ``peak chasing''. The primary information inputs to peak chasing involve a cognitive parameter called ``Functional Acuity'' (FA), and parameters related to the ability to respond to changes efficiently, which we collectively call ``Functional Operability'' (FO). 

\item[]Phenomena at the \textbf{Meso-Cognitive scale} involve reasoning and decision-making related to whether to continue or abort the current measurement. 

\item[]Phenomena at the \textbf{Macro-Cognitive scale} involve reasoning and decision-making related to the course of the experiment, especially regarding how long to gather data on a given sample, and in what order samples should be run. 
\end{itemize}

\clearpage
\subsection{Micro-Cognitive: Peak Chasing, Functional Acuity, and Functional Operability} \label{peak-chasing}

In the micro-cognitive domain we model an abstract task that we call ``peak chasing'', a common type of optimization task, akin to balancing a broom on one's hand. At LCLS peak chasing involves actuating instrument controls to try to optimize measurement SNR. There are hundreds --- perhaps thousands --- of parameters that control the linac$\rightarrow$XFEL$\rightarrow$instrument system. In some cases the ``peak'' being ``chased'' is literally a peak in a histogram or similar graph, or a single value on a display. Sometimes it is a statistic reported by the Real Time Data Analyst. And sometimes the ``peak'' is more subtle, such as a ``correct looking'' display. In our model the peak being chased is the alignment of two values, which one can think of as representing the horizontal position of a stream of liquid versus the horizontal position of the beam. The quality of the data (SNR) collected depends upon how close the beam is to the stream. The liquid undergoes a one-dimensional random walk. The instrument operator must first notice that the beam and stream are not aligned, and then move the beam left (X-) or right (X+) to ``chase'' the stream. 

The operator's ability to carry out this task is controlled by two abstract parameters (that is, implemented by one or more specific model parameters): Functional Acuity (FA) and Functional Operability (FO). FA is the operator's (or, more generally, the team's) ability to \textit{recognize} the need to take an action. FO is their ability to actually \textit{take} the appropriate action --- more precisely, how long it takes to take the appropriate action (controlled by the Noticing Delay (ND) parameter, among others). These are directly related to the important resources of time and attention discussed in section \ref{resources}. 

The term ``functional'' emphasizes that these are not purely cognitive concepts, but are functions of the interaction between individual cognition, team communications, and technical displays and controls. Functional Acuity and Functional Operability have a temporal as well as informational component. For example, in addition to a parameter that controls how far the beam and stream have to be before the operator can notice that an action must be taken (literally the acuity), the noticing delay slows the operator's response. Also, if the operator has to change which button is being pressed, e.g., to switch the direction a motor is being stepped, a cost is incurred based on the distance from the current button to the new one. 

Both Functional Acuity and Functional Operability can be thought of as UI-related parameters. However, differences in Functional Acuity arise from the interaction between cognition and the \textit{display} characteristics of the instrument UI, whereas differences in Functional Operability arise from the interaction between cognition and the \textit{control} characteristics of the instrument UI. Whereas changes to the displays or controls would usually affect only one or the other of these, more general interventions, such as the operator's alertness, experience with the system, or changes in the operating protocols, could affect either or both of these parameters, as well as other aspects of the model, such as error rates.

\subsection{Meso-Cognitive: When to End a Measurement}

As previously noted, when to end a measurement, that is when to stop taking data on a particular sample, is among the most important decisions made by the team. In reality, many considerations go into this decision, including the stability of the beam, the apparent quality of the data per secondary data processing, the quality of the data that has been obtained so far, the amount of time remaining in the day or entire run, how much sample remains, and considerations relating to the overall experiment such as the importance of the sample, and possibly tertiary data analysis that may have been completed and can provide complete analysis of previous measurements. At this scale our model considers only the current rate of improvement in SNR, and the beam time remaining. 

\subsection{Macro-Cognitive: Sample Planning and Re-Planning}
\label{macro}

The dynamic environment of an LCLS experiment is markedly different from puzzle-like problem-solving where one usually has the luxury of time and a small set of possible actions. We previously likened the measurement process as akin to balancing a broom stick on one's hand,  but it is made even more complicated by the nature of the novel physics being studied at LCLS, and by a constant barrage of tiny (and sometimes large) problems, requiring real-time re-planning.\cite{Hayes-Roth1995} While some adjustments, like shortening a measurement if one is obtaining all the needed high-quality data, are relatively straightforward, others, such as extending a measurement, or even bringing in additional materials from off-site if necessary, can be more challenging given the strictly limited time frame. In our model, the only information used to make these decisions is the information available regarding how well the measurements have been going so far, how much beam time remains, and the number and quality of samples remaining to test. 

\subsection{Measurement Error and Performance Quality of \linebreak \mbox{Samples}} \label{tepq}

Central to the model (and, of course, to the real LCLS setting) is the quality of the data being gathered --- SNR (\S \ref{expmes}). How SNR is defined depends upon the details of the kind of data being gathered. Regardless, it always involves some version of the standard error, i.e., the standard deviation of the data divided by the square root of the number of data points. Therefore, we simply use standard error as a stand-in for SNR, and model the goal of a given measurement as being to reach a particular target standard error, which we call ``Target Error'' (TE), and which is 0.001 by default. 

Samples in our model only differ from one another in terms of a single parameter, called ``Performance Quality'' (PQ), related to the types of samples (\S \ref{expmes}). A sample's PQ represents how easy it is to get data from this sample, based upon either prior experiments, for example in a synchrotron, or theoretical calculation. Higher PQ samples will produce higher quality data, that is, they will reach TE with fewer data events than samples with lower PQ. 

Counter-intuitively, more important samples have lower PQ. The reason that lower PQ samples are considered more important is that if it was easy to get data from a given sample (high PQ) then one would have done so already, or one would be sure of their theoretical prediction. Thus, it is \textit{less} important to get high quality data at LCLS from samples that have \textit{high} PQ. Moreover, the quantity of low PQ (more important) samples on hand is also often limited --- again, if samples were unlimited one would have been able to get adequate data on these in a less time-constrained setting such as a synchrotron. One does not want to be wasting rare and important low PQ samples, especially early in the experiment, when the SNR is low for extrinsic reasons, such as overall system instability or operator inexperience with this particular experiment. Therefore, one generally will want to run high PQ (less important) samples early on until one has high confidence in the data coming from the instrument. Put another way, high PQ (less important) samples are those for which you essentially know what data you are likely to get, whereas low PQ (more important) samples are ones that are harder to get data from in either (or both) of theoretical modeling or previous studies (e.g., in a synchrotron). The whole reason for coming to LCLS and running on the XFEL is that it is hard to get data on the low PQ (more important) samples. 

\subsection{Error Adjustment and Cutoff Time}

At LCLS, where both time and samples are limited, it may be difficult to reach the desired Target Error (TE). As discussed above, one wants to run low importance, high PQ, samples early, during instrument stabilization, and more important, low PQ, samples later, but not so late that the team will run out of beam time. Although real experiment managers might dynamically decide the order in which to measure samples, in the present paper we always model samples as run in High-PQ (less important)$\rightarrow$Low-PQ (more important) order. The simulated experiment manager in our model can decide what TE to reach for, and when to cut off a given measurement. The model can also be constrained regarding the total beam time (``Cutoff-Time''), where ``True'' (yes) means that the experiment will be stopped after an allotted time, and may be allowed to adjust the TE (``Adjust-Error''), where ``True'' (yes) allows the experiment manager to change the TE. In the most realistic case Cutoff-Time and Adjust-Error are both True. In this case the team may adaptively decide to shoot for a higher or lower TE, depending upon how the measurements are going and the amount of time remaining, but at the end of their allotted time the experiment is halted.  

\section{Example Results}

In this section we describe some example results obtained from our model that demonstrate various ways that it could be used to investigate changes in the user interfaces or workflows of LCLS instrument-side experiments. These are discussed in more detail in the second of the three supplementary videos. In each case the vertical axis records the number of events (data points) required to reach the Target Error (default=0.001), over Performance Quality (PQ on the horizontal axis). As mentioned just above, all simulation runs presented here start with the lower importance, higher PQ samples, and proceed to the higher importance, lower PQ samples, that is reading right to left on the graphs below.

\begin{figure}[!htbp]
    \centering
      \makebox[\textwidth]{\includegraphics[width=\textwidth]{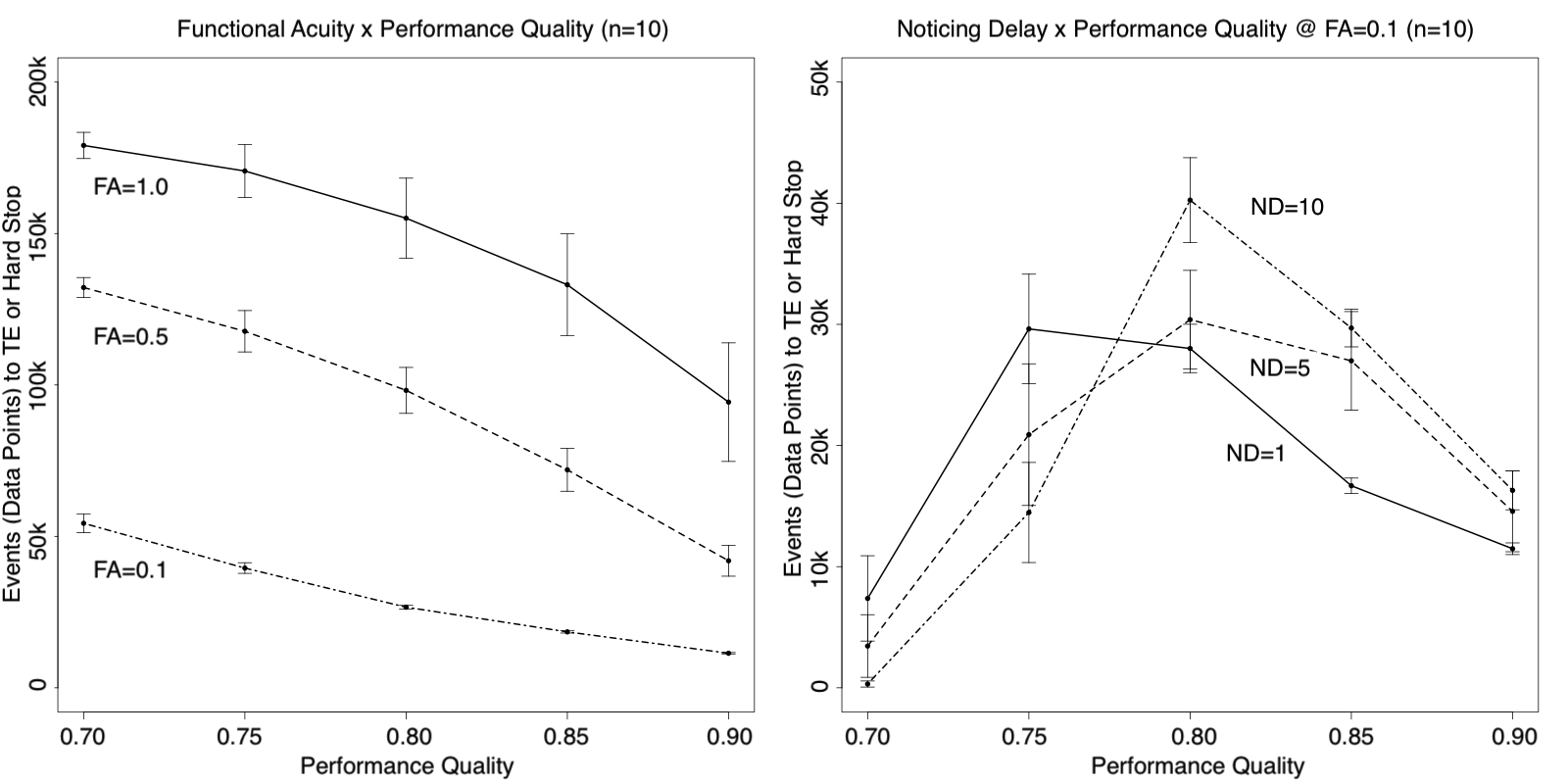}}
        \caption{Left: Varying Functional Acuity (FA = [0.1, 0.5, 1.0]) with Adjust-Error=False and Cutoff-time=False. Lower FA corresponds to noticing finer misalignments, thus requiring fewer samples to reach TE. Runs start at high Performance Quality (PQ) and go to progressively lower PQs --- that is, from lower importance samples to higher, or \textit{right to left}. With Adjust-Error=False and Cutoff-time=False the experiment is allowed to run to reach TE=0.001 because the beam time never expires. Right: Fixed FA = 0.1, varying Noticing Delay (ND = [1, 5, 10]) with Adjust-Error=True and Cutoff-time=True. Higher ND corresponds to slower response to misalignments. ND has much less differential effect at these levels (cf. Figs. \ref{fig8}-\ref{fig9}). With Adjust-Error=True and Cutoff-time=True some data is gathered on all samples even though the team's beam time expires (cf. Figure \ref{fig9} left), although not as much data as if they are allowed to continue, as on the left.}
    \label{fig7}
\end{figure}

The simulation results in Figure \ref{fig7}-Left scan three values of Functional Acuity (FA), with Adjust-Error=False and Cutoff-time=False, thus allowing the team to remain as long as necessary to reach the default Target Error (TE) of 0.001. Unsurprisingly, it requires less data to reach TE 0.001 as PQ increases (right to left), and as FA improves (becomes smaller), so that the operator notices that the beam and stream are not aligned at smaller distances of misalignment and can respond more quickly. The results depicted in Figure \ref{fig7}-Right locks FA at 0.1 (the best performing value from \ref{fig7}-Left), and compares Noticing Delays (NDs). In this case, Adjust-Error=True and Cutoff-time=True, which is the most realistic case: The team is kicked off the machine at a fixed time, but they are allowed to adjust TE to try to get some data on all samples. At first glance Noticing Delay makes no apparent difference; All the values are more or less the same. However, as PQ increases, the amount of data gathered appears to be smaller at both high and low PQ, but higher at mid values of PQ. This happens because the simulated experiment manager observes that they are going to run out of time with the default TE, and adjusts the TE upward so that some data is obtained for even the last/most important/lowest PQ samples, although they eventually run out of time anyway. More flexible re-planning would improve the data quality for these most important (lowest PQ) samples. 

\begin{figure}[!htbp]
    \centering
      \makebox[\textwidth]{\includegraphics[width=\textwidth]{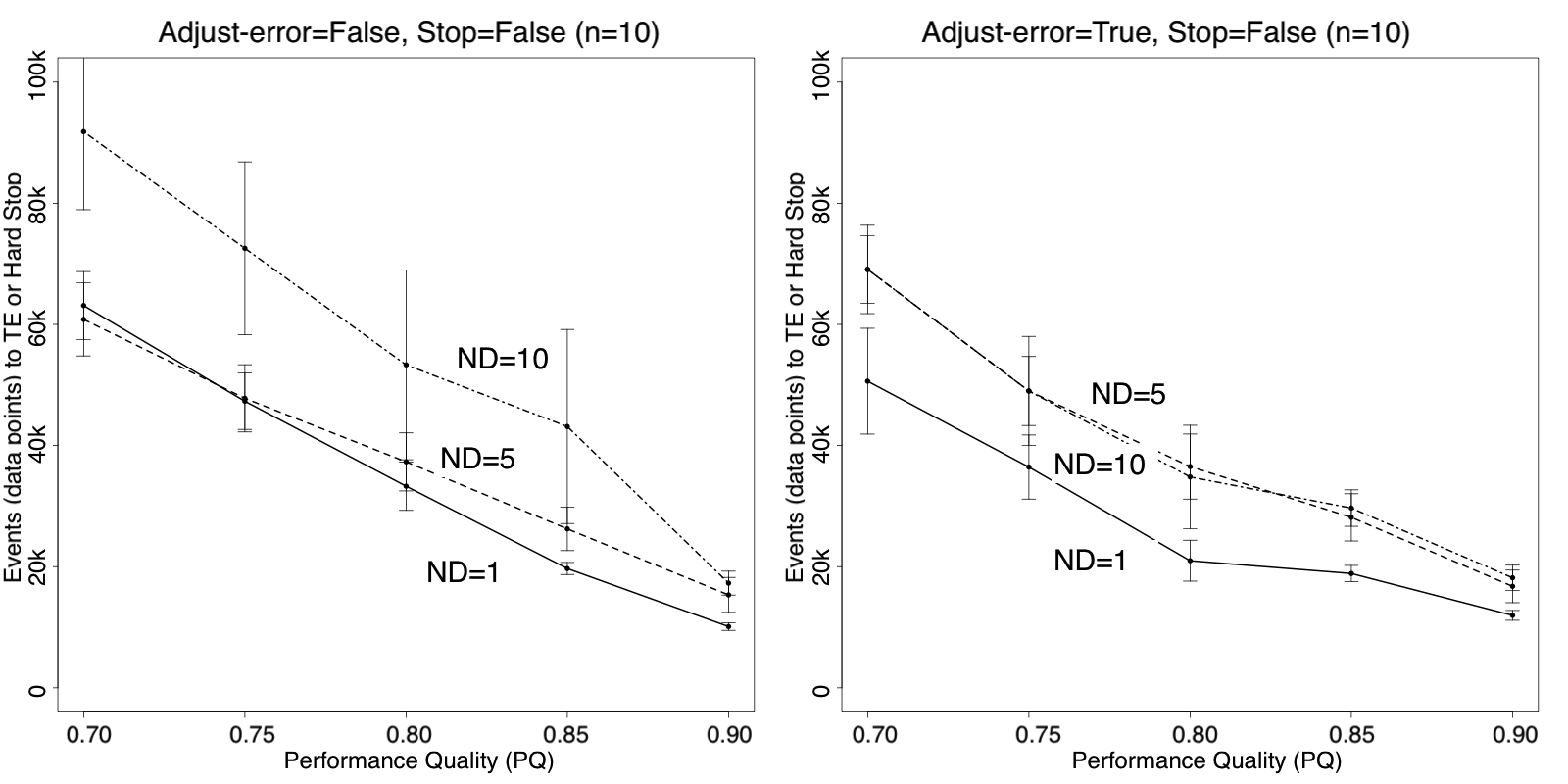}}
        \caption{Cutoff Time (Stop) = True across Noticing Delay (ND) = [1, 5, 10] and Adjust-Error = [False, True].}
    \label{fig8}
\end{figure}

\begin{figure}[!htbp]
    \centering
      \makebox[\textwidth]{\includegraphics[width=\textwidth]{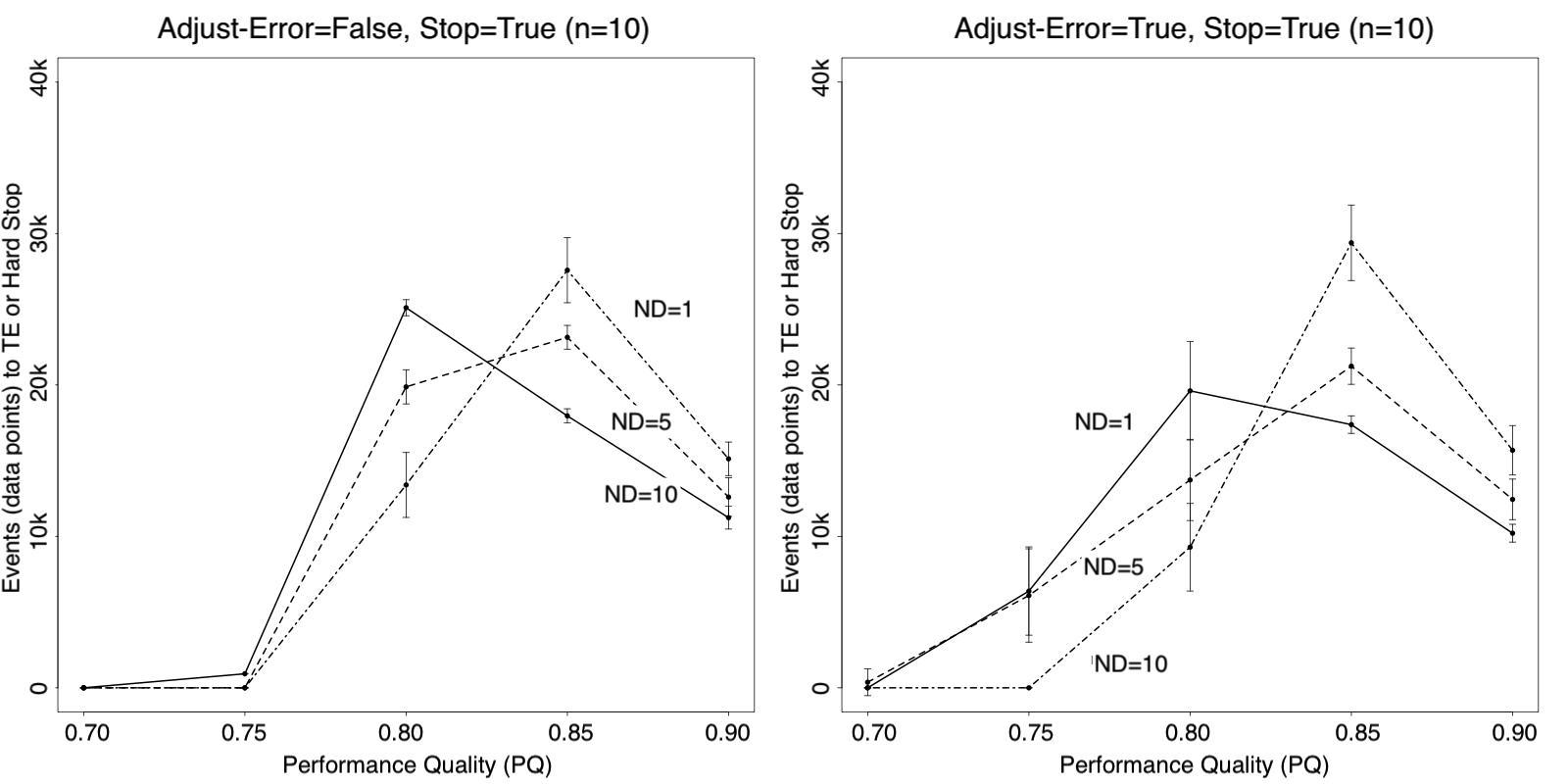}}
        \caption{Cutoff Time (Stop) = False across Noticing Delay (ND) = [1, 5, 10] and Adjust-Error = [False, True].}
    \label{fig9}
\end{figure}

\newpage
Figures \ref{fig8}-\ref{fig9} depict four explorations with the model, scanning Noticing Delay (ND) at three values, across all combinations of the conditions Adjust-Error and Cutoff Time (Stop). The goal of the team in all cases is to achieve a good (low) error rate within the allotted time. As above, TE is initially 0.001 (the default). With Adjust-Error=False this never changes. However, with Adjust-Error=True the experiment manager will adjust the TE based on the time remaining and the observed performance per unit Performance Quality (PQ) to this point. (These algorithms are reviewed in the supplementary videos.) All of the present runs go from higher to lower PQ, that is right to left. 

The first (rightmost) sample TE will always be 0.001, reagrdless of the value of Adjust-Error. When the team is able to adjust the error rate (Adjust-Error=True, right of Figures \ref{fig8}-\ref{fig9}), the second through last (fifth) sample's TE can be adjusted to try to stay within the given time. When the time is strictly limited Cutoff-time(Stop)=True (Figure \ref{fig9}) the team is eventually kicked off the machine. The most realistic condition, is Figure \ref{fig9}-Right: Adjust-Error=True, Stop=True, and one can see by comparing the amount of data gathered on the last two (leftmost) sample runs between the left and right of Figure \ref{fig9}, that when the team is allowed to adjust the error target (\ref{fig9}-Right), they gather some, although not very much, data on the last two (leftmost) critical samples, whereas if they cannot adjust error (left side of both figures), but are strictly time-limited (Figure \ref{fig9}-Left) essentially no data is gathered on the last two (leftmost) critical samples because the experiment is forced to end short of time. For comparison Figure \ref{fig8}, with no time limitation (Stop=False), is an unrealistic situation but tells us how much data needs to be collected to reach a given TE. Figure \ref{fig8}-Left (Adjust-Error=False, Stop=False) tells us how long it will take to reach TE = 0.001, whereas the \ref{fig8}-Right (Adjust-Error=True, Stop=False) tells us how long it will take to reach an adjusted TE. The effect of ND is more pronounced in Figure \ref{fig8} because they run significantly longer than those in Figure \ref{fig9}. These results are also discussed in the second supplementary video.

\newpage

\newpage

\section{Discussion}

The model described in this paper is the culmination of several years of observational research at LCLS, during which our team interacted with numerous LCLS staff and external scientists. As we became regular observers during experiment operations, and sat in on experiment planning and engineering meetings, we were sometimes looked-to to provide guidance regarding control and display design decisions. Usually we demurred, having only qualitative data, and so weak confidence in our ability to provide useful guidance. The impetus for developing the model described in this paper was to put such guidance on a firmer, quantitative footing.

Although the importance of complex team cognition is widely recognized, and has been studied in detail in many settings (e.g., the notable studies of naval and aircraft operations by Hutchins and coworkers\cite{Hutchins2013,Hutchins2014}), most computational cognitive models focus on individual cognition. Among the nearly 900 publications listed in the ACT-R publication repository\cite{ACT-R}, only 11 are tagged with ``Communication, Negotiation, and Group Decision Making'', and these usually model very simple  settings (e.g., Matessa, 2001\cite{Matessa2001}). Another line of work (e.g., Goldstone and Janssen, 2005\cite{GoldstoneJanssen2005}) has pursued multi-agent models with many nearly identical simple agents, often with the ability to learn. That work focuses on the evolution or self-organization of differentiated agents and/or complex communities. Complex teamwork has been computationally modeled in business and military settings (e.g., Cooke and Myers, 2009\cite{CookeMyers2009}, Ball et al, 2010\cite{syntheticteammate}) but far more rarely in scientific settings. Whereas scientific teamwork has been studied (e.g., Traweek, 1988\cite{Traweek88}) and scientific reasoning computationally modeled (e.g., Shrager and Langley, 1990\cite{sl90}) we have not found computational models of scientific teamwork designed to simultaneously capture both individual cognition and collaboration. Not only does our model attempt to capture these complexities in the unique setting of a scientific user facility, but also captures aspects of planning and opportunistic re-planning specific to LCLS. 

Scientific user facilities, such as LCLS, are particularly cognitively interesting and difficult precisely because they are on the cutting edge of both science and human performance. There are, by definition, no perfect simulators for state-of-the-art science; If there were, one would not need real instruments! And although the instrument scientists are expert at their instruments, they are not expert at each new experiment. Moreover, they are usually working with external users who have widely varying skills, knowledge, and experience, and with state-of-the-art instruments that are operating at the bleeding edge of measurement capability and stability.

Even though our current model is only a rough analog to the highly complex LCLS setting, it produces sensible results across a variety of parameter ranges, and provides insight into how these various factors and scales of cognition interact to impact the efficiency of experimental operations at LCLS. Improving Functional Acuity (FA), which can arise from UI and workflow changes, can produce faster response to beam misalignments, leading to increased overall experimental efficiency. In contrast, improved Noticing Delay (ND) produces less significant impacts on efficiency, except in unusual operating regimes, for example when there is no time limit. We also investigated quality and duration trade-offs by adjusting whether the Target Error (our stand in for SNR) is dynamic or fixed, and by imposing cutoff times, or allowing runs to continue indefinitely. Unsurprisingly, fixed target error requirements produce higher fidelity given unlimited time. Allowing the Target Error to adapt yields better time management, producing higher overall signal on important samples in the more realistic condition where measurements are truncated by limited beam time. 

These observations showcase how operational efficiency depends substantially on factors related to the human–machine interface such as Functional Acuity, its operation such as Functional Operability, and cognitive factors such as attentiveness, amount of experience, and fatigue (represented by Noticing Delay and other parameters). Enhancing these aspects of the LCLS setting --- by training, redesign of controls and displays, and workflow changes --- could substantially increase scientific productivity.

Our model merely scratches the surface of the potential of model-based cognitive engineering efforts as applied to scientific user facilities such as LCLS. Although we have described only a handful of computational experiments with the model, it is capable of simulating a much wider range of parameters. Moreover, relatively simple extensions could easily broaden its scope to model specific displays, controls, and workflows, demonstrating how changes in these might impact efficiency at all scales. The model should also be extended to capture more inter- and intra-team communications. Elaborations such as those would also permit the model to be validated with respect to detailed measurements of individual and team behavior. Most importantly, and perhaps most difficult to implement, models of learning at all scales should be introduced in order to capture  short term individual learning such as the instrument scientists becoming familiar with each new experiment; social learning such as the team learning to work together; and longer term learning such as expanded staff training.

As science probes ever more deeply into the complexities of nature, large scale, beyond-the-state-of-the-art efforts like those at LCLS, and other scientific user facilities, are rapidly increasing in importance. These facilities are in constant flux at every scale. Efforts such as ours that attempt to model the system at multiple scales should be helpful in increasing the efficiency of the science carried out at these sites by helping to optimize the human-in-the-loop workflows that will be a significant component of the operations of scientific user facilities for the foreseeable future. 

\pagebreak

\section{Supplementary Material}

Three supplementary videos\cite{vids} provide much more detail on all of the above. The first is a conceptual overview. The second demonstrates the model, including discussion of the results depicted in Figures \ref{fig7}-\ref{fig9}. The third reviews the algorithms and code, including walk-throughs of important code segments. The code is open source and available on github.\cite{githubmodel}
 
\section{Acknowledgments}

The research described here arises from a project begun in 2019 by Devangi Vivrekar and Paul H. Fuoss, with guidance from Jeff Shrager. The research was continued by Jeff, Paul, and Teddy Rendahl, and then by Jeff, Paul, Wan-Lin Hu, and Jonathan Segal. The LCLS and SLAC staff and visiting scientists that permitted us to observe them working, and sat for interviews over those years, are too numerous to mention, but we are especially grateful to Andy Aquila, Mark Hunter, and Meng Liang who put up with us observing their experiments (and asking naive questions) on multiple occasional, and especially to Dan Flath and Jana Thayer for valuable guidance, support, and encouragement. We thank Nick Briggs and several anonymous reviewers for greatly improving this paper. Figures 3 and 5 are reprinted with permission from the SLAC National Accelerator Laboratory. Use of the Linac Coherent Light Source (LCLS), SLAC National Accelerator Laboratory, is supported by the U.S. Department of Energy, Office of Science, Office of Basic Energy Sciences under Contract No. DE-AC02-76SF00515.

\section{Contributions of Authors}

Jonathan Segal did most of the model coding, ran the simulations, and created the graphics. Wan-Lin Hu did most of the field data collection and analysis. Frank Ritter provided important guidance on cognitive engineering and human factors. Paul H. Fuoss conceived the project, obtained the funding, dealt with all administrative aspects, made most of the personnel connections, and was our primary applied physics informant. Jeff Shrager managed the project, participated in the observations and model coding, wrote this paper, and created the supplementary videos. All authors reviewed and commented on the paper and videos.

\pagebreak

\bibliographystyle{unsrt}
\bibliography{LCLSpaper}
\end{document}